\documentstyle[12pt, epsf]{article}

\def\dalemb#1#2{{\vbox{\hrule height .#2pt
        \hbox{\vrule width.#2pt height#1pt \kern#1pt
                \vrule width.#2pt}
        \hrule height.#2pt}}}

\let\a=\alpha   \let\d=\delta \let\e=\epsilon
  \let\th=\theta  
 \let\m=\mu \let\n=\nu  \let\p=\pi 
\let\s=\sigma \let\t=\tau    
 
       \let\D=\Delta \let\Th=\Theta

\def\nn{\nonumber} \def\bd{\begin{document}} \def\ed{\end{document}}
\def\ds{\documentstyle} \let\fr=\frac \let\bl=\bigl \let\br=\bigr
\let\Br=\Bigr \let\Bl=\Bigl
\let\bm=\bibitem
\let\na=\nabla
\def\tU{{\widetilde U}}
\let\pa=\partial \let\ov=\overline
\def\ie{{\it i.e.\ }}
\newcommand{\be}{\begin{equation}}
\newcommand{\ee}{\end{equation}}
\def\ba{\begin{array}}
\def\ea{\end{array}}
\def\ft#1#2{{\textstyle{{\scriptstyle #1}\over {\scriptstyle #2}}}}
\def\fft#1#2{{#1 \over #2}}
\def\F#1#2{{ F_{#1}^{(#2)} }}
\def\cF#1#2{{ {\cal F}_{#1}^{(#2)} }}

\def\R{{\bf R}}
\def\sst#1{{\scriptscriptstyle #1}}
\def\oneone{\rlap 1\mkern4mu{\rm l}}
\def\e7{E_{7(+7)}}
\def\td{\tilde}
\def\wtd{\widetilde}
\def\im{{\rm i}}
\def\bog{Bogomol'nyi\ }
\newcommand{\ho}[1]{$\, ^{#1}$}
\newcommand{\hoch}[1]{$\, ^{#1}$}
\newcommand{\bea}{\begin{eqnarray}}
\newcommand{\eea}{\end{eqnarray}}
\newcommand{\ra}{\rightarrow}
\newcommand{\lra}{\longrightarrow}
\newcommand{\Lra}{\Leftrightarrow}
\newcommand{\ap}{\alpha^\prime}
\newcommand{\bp}{\tilde \beta^\prime}
\newcommand{\cB}{{\cal B}}
\newcommand{\cO}{{\cal O}}
\newcommand{\vecx}{\vec{x}}
\newcommand{\vecy}{\vec{y}}
\newcommand{\vecp}{\vec{p}}
\newcommand{\vecq}{\vec{q}}
\newcommand{\tr}{{\rm tr} }
\newcommand{\Tr}{{\rm Tr} }
\newcommand{\NP}{Nucl. Phys. }

\def\sst#1{{\scriptscriptstyle #1}}
\def\0{{\sst{(0)}}}
\def\1{{\sst{(1)}}}
\def\2{{\sst{(2)}}}
\def\3{{\sst{(3)}}}
\def\4{{\sst{(4)}}}
\def\5{{\sst{(5)}}}
\def\6{{\sst{(6)}}}
\def\7{{\sst{(7)}}}
\def\8{{\sst{(8)}}}
\def\ve{\varepsilon}
\def\vf{\varphi}
\def\F{\Phi}
\def\e{\epsilon}
\def\wg{\wedge}

\newcommand{\tamphys}{\it Center for Theoretical Physics,\\
Texas A\&M University, \\College Station, Texas 77843}

\newcommand{\auth}{I.Y. Park}

\def\thb{\bar{\theta}}
\def\Thb{\bar{\Theta}}
\def\barp{\bar{p}}
\def\barq{\bar{q}}
\def\barc{\bar{c}}
\def\bard{\bar{d}}
\def\e{\epsilon}

\def\th{\theta}
\def\Th{\Theta}
\def\vth{\vartheta}
\def\btheta{{\bar\theta}}
\def\ttheta{{{\tilde\theta}}}
\def\bttheta{{{\bar\ttheta}}}
\def\vth{\vartheta}

\def\ra{\rightarrow}
\def\N{{\cal N}}
\def\F{{\cal F}}
\def\uM{\underline{M}}
\def\uN{\underline{N}}
\def\uP{\underline{P}}
\def\cc{\circ}
\def\eqv{\equiv}

\def\ni{\noindent}

\def\Ep{E^{{}^{(+)}}}
\def\Em{E^{{}^{(-)}}}

\def\Mp{M^{{}^{(+)}}}
\def\Mm{M^{{}^{(-)}}}

\begin{document}
\begin{flushright}
\hfill{CTP TAMU-21/01}\\
\hfill{hep-th/0106078}\\
\hfill{}\\
\end{flushright}

\vspace{20pt}

\begin{center}
{\large {\bf  Strong Coupling Limit of Open String: Born-Infeld Analysis}}

\vspace{30pt}
\auth

{\tamphys}     

\vspace{40pt}

\begin{abstract}
We consider a large coupling limit of a Born-Infeld action in a
curved background of an arbitrary metric and a two form
field. Following hep-th/0009061, we go to the Hamiltonian
description. The Hamiltonian can be dualized and the dual action
admits a string-like configuration as its solution. We interpret
it as a {\em closed} string configuration. The procedure can be
viewed as a novel way of bringing out the appropriate degrees of
freedom, a closed string, for a open string under the strong
coupling limit. We argue that this interpretation implies a large
number of dual pairs of gauge and gravity theories whose particular
examples are AdS/CFT and matrix theory conjectures.

\end{abstract}

\end{center}

\newpage

\section{Introduction}
The fact that opens strings have end points gave rise to many
interesting results in the recent progress of string theory.
Most notably it led to the discovery of D-branes \cite{pol}.
It is plausible that there are yet to be discovered physics
associated with the end points, some of which could be
important. In the studies of D-branes it has been fruitful to consider
extreme values, such as zero or
infinity, of various parameters of the system under
consideration. This is, for  example, what one does in AdS/CFT
correspondence  \cite{mal,gkp,w1}, a {\em large} N duality between
gauge theory and gravity theory.

In this note we consider the dynamics of the end points of an open
string by associating a ``quark'' pair with them. In particular
we consider a strong coupling limit where the string coupling
constant approaches {\em infinity}. Then it seems natural, at least at a
naive level, to expect that the two end points get ``stuck together'', which
will in turn suggest a novel mechanism in which the original open string
may be converted into a closed string. It is the aim of this paper to
investigate this possibility and its implications.

Part of the motivation of this work came from \cite{iyp} (see
\cite{kv,dgks,smol} for related discussions) where it was observed that
the gauge/supergravity duality may in fact be deduced as a low
energy limit of a ``duality'' between two different {\em stringy}
descriptions of D-branes, open and closed. There, the ``duality''
was taken to be a starting point as an axiomatic assumption. Given
that we have an open string description on one side and a closed
string description on the other side, a conversion mechanism is
likely to be relevant for the understanding of the relation
between the two descriptions. Here we propose that the axiomatic
duality should be associated with a strong-weak duality of an open
sting in the sense we will discuss in the later part of this note.

In the first part, we are concerned with how one might see the
conversion of open strings into closed strings. It would be ideal
if this picture could be realized quantitatively at the level of
full string theory. However, given the limitations of the full
string theory techniques for a strongly coupled system, it will be
easier to turn to the low energy effective action of open strings,
a Born-Infeld (BI) action.

The simplicity gained by resorting to the effective action does
not come without a price: one has to face the issue of justifying
the results because, as
well known, a Born-Infeld type action has its limitations
(For reviews of a BI action see \cite{tsey,gib}.) We discuss some of these
issues in due course.

If the physics of the strong coupling limits indeed converts open
strings into closed strings, it should be a general phenomenon
that could and should happen to a generic open string system. Long
ago the authors of \cite{no} considered a bosonic Born-Infeld
action. (See \cite{lv,lz} for more recent discussions of a strong
coupling limit.) They argued that the Lagrangian in the strongly
coupled limit admits a solution that provides a simple description
for {\em closed} strings. The dynamical generation of closed
strings have also appeared in \cite{ghy,s3,kls} in the tachyonic
context. (see also \cite{sh,ckl} and \cite{hkm,kmm}.) In
particular it was the U(1) confinement mechanism \cite{yi,bhy}
that is responsible for the appearance of closed strings in the
work of \cite{ghy}. For our purpose it is intriguing to note that
all the mathematical manipulations of \cite{ghy} carry over when
we replace the tachyon potential, V(T), by the usual tension,
$\tau$.  The limit, $V\ra 0$, can be viewed as to correspond to
the strong coupling limit where the tension vanishes, $\tau\ra 0$.

The fact that the manipulations remain the same (other than
$V\ra\tau)$ strengthens our belief that a confinement
mechanism should, in fact, be a general feature of open string
systems, not restricted to a tachyonic system. Therefore one
should be able to see the same feature for a BI action in a more
general background such as a curved one or a background with a
B-field\footnote{Various BI actions in a curved background with 
a (non-) constant B-field 
were previously considered in \cite{koji}.}. We will see
that this is true. The necessary
manipulations is a generalization of the steps presented in
\cite{ghy}.

Although our calculations are limited to a low energy limit, we
view the results as evidence for the ``open-closed string
duality'' and study its implications. In particular we note that
AdS/CFT may be the low energy realization of the duality. We also
argue that the duality may explain the matrix theory conjectures.

The paper is organized as follows: In section 2, a bosonic
BI-action is considered in a background of a general metric
$g_{\m\n}$ and a constant $B$-field. We go to the Hamiltonian
formulation.
After rewriting the Hamiltonian in terms of the canonical
variables, we dualize it to a new Lagrangian. The solution of the
dual action has its support along the two dimensional surface that
can be viewed as a string world sheet. We interpret the solution
as a closed string configuration. Substitution of the solution
into the dual action yields a Nambu-Goto type action in the {\em
same} curved background. We discuss, in section 3, the
implications of our results for AdS/CFT and matrix theory
conjectures. Section 4 contains the conclusions with open problems.

\section{Dual description of Born-Infeld Hamiltonian}

Here we generalize the calculations in the section 4 of
\cite{ghy}. Consider a BI Lagrangian in the presence of a metric,
$g_{\m\n}$, and a constant\footnote{Or one could consider
a non-constant, {\em closed} two form field as in \cite{koji}.}  
two form field, $B_{\m\n}$,
\bea {\cal L} &=& -\;\tau \sqrt{-\mbox{Det}\left(g_{\m\n}+M_{\m\n}
            \right) } \label{bi} \eea
where we have introduced a shorthand notation,
$M_{\m\n} \!\!\equiv \!
F_{\m\n}+B_{\m\n}$. For simplicity we impose the following
conditions on the metric,
\be g_{\cc i}=0=g_{i \cc} \ee
The full discussion is presented in the appendix.
The determinant can be rewritten as
\be -\mbox{Det}\left(g_{\m\n}+M_{\m\n}
            \right)= -\mbox{Det}( h_{ik} )\,g_{\cc \cc} - M_i
D^{ik} M_k \ee
where $M_i \eqv M_{\cc i}$.
Above we have introduced similar notations as those in \cite{ghy},
\bea 
h_{ij} = g_{ij}+M_{ij}\;\;\;,\;\;\;
D^{ij} = (-1)^{i+j}\D^{ji}(h) \;\;\;,\;\;\;
D=\mbox{Det}(h)\,  h^{-1}
 \eea
where $\D$ represents the minor matrix of
$h$. Since the $B$-field is a background we consider the canonical
momenta of $E_i$ but not of $M_{\cc i}$. The Hamiltonian is then
\bea {\cal H}_B = \pi^i E_i-{\cal L}= \pi^i M_i-\pi^i B_i -{\cal
L} \eea
where $B_i\eqv B_{\cc i}$. The canonical momenta are
\be \pi^i = \frac{\delta {\cal L}}{\delta \dot{A_i}}
      =\fr{\t }{   \sqrt{  -\mbox{Det}( g_{\m\n}+M_{\m\n} )  }   }\;
             \fr{1}{2} ( M_k D^{ki} + D^{ik}M_k )  \nn\\
\ee
After some algebra one can show
\bea {\cal H}_B  &=& -  \fr{\t\,g_{\cc \cc}\,\mbox{Det}(h)}
                {\sqrt{-\mbox{Det}(g_{\m\n}+M_{\m\n})} } -\pi^i B_i \nn\\
            &=&    \sqrt{ -g_{\cc \cc} \left(\pi^i g_{ij} \pi^j\!
                   +\!\left[(F_{ij}+B_{ij})\pi^j
                   \right]^2 + \t^2\mbox{Det}(h)\right)
                   \!} \;-\pi^i B_i
\eea
In the strong coupling limit, one can drop the last term inside
the square root. Performing a Legendre transformation \cite{tse},
\bea {\cal L}_B' &=& {\cal H}_B-\frac{1}{2} F_{ij}K^{ij}
         \label{duality}
\eea with
\bea K^{ij} = 2\frac{\delta {\cal H}}{\delta  F_{ij}}
       =   -\frac{g_{\cc \cc}}{\cal H}\,
                  \left( {M^i}_k\pi^{k}\pi^{j}- {M^j}_k\pi^{k}\pi^{i}
                   \right)
\label{K} \eea
leads to
\bea {\cal L}_B' &=& {\cal H}_B-\frac{1}{2} M_{ij}K^{ij}
                                     +\fr12 B_{ij}K^{ij}          \nn\\
            &=& \sqrt{-{g_{\cc \cc}}\pi^2 - K^{ij}K_{ij}/2}-\pi^i B_i
                            +\fr{1}{2} B_{ij}K^{ij}
\label{lpb} \eea
Therefore eq (\ref{bi}) admits a compact dual description
\be {\cal L}'=\sqrt{-\fr12{\cal K}^{^{\m\n}}{\cal K}_{_{\m\n}}}
                    +\fr{1}{2}B_{\m\n}{\cal K}^{\m\n}
\label{Kappa-L} \ee
where
\bea {\cal K} &=&-{g_{\cc \cc}}\pi_{i}dt\wedge dx^{i}
              +\fr{1}{2}K_{ij}dx^{i}dx^{j} \label{Kappa}
\eea
\ni The same equations that were satisfied by ${\cal K}$ in the
flat case \cite{ghy} are also satisfied here: from the definition
of ${\cal K}$ it is easy to show that it satisfies a constraint
${\cal K}\wedge {\cal K}=0$. The Bianchi identity, $d F=0$, is now
translated into the equation of motion of ${\cal K}$,
\be 0=d \left(\frac{{\cal K}}{\sqrt{-{\cal K}^2/2}}\right),
\label{eom} \ee
There is another constraint equation that ${\cal K}$ must satisfy:
\be
\partial_{\m}{\cal K}^{\m\n}=0
\label{b} \ee
This corresponds to the equation of motion and the Gauss
constraint of the original description.\footnote{The equation of
motion (\ref{eom}) admits a scaling symmetry ${\cal K}\rightarrow
f{\cal K}\;,\; \mu \rightarrow \mu /f \label{scale}\; $ where $f$
is an arbitrary function. $f$ should be a constant for the similar
reason discussed in \cite{no}.} With these conditions one can
write down the following solution,
\be {\cal K}^{\m\n}=\int \d(X-Z(\s))\;dZ^\m \wedge dZ^\n
\label{ss} \ee
Substitution of (\ref{ss}) into (\ref{Kappa-L}) gives a Nambu-Goto
type action in the same background,
\be S=\int d^2\s\; \sqrt{\mbox{Det}(\pa_a X^\m \pa_b X^\n
g_{\m\n}) }
              + \fr{1}{2}\int d^2 \s \e^{ab} \pa_a X^\m \pa_b X^\n B_{\m\n}
\label{ng}
\ee
We view (\ref{ng}) as a {\em closed} string action.

\section{Interpretation and Implications}
We have considered a Born-Infeld action and its strong coupling
limit. Following the literature, we went to the Hamiltonian
formulation to study the physics of the strong coupling limit. The
Hamiltonian can be dualized to yield an action that allows a
connection to a Nambu-Goto type string action, (\ref{ng}). 
We interpret this as a low energy realization of the conversion of an open 
string into a closed string.

As we discussed in the introduction, the transition of an open
string into a closed string should be a general phenomenon.
Therefore it should be possible to extend the results to
supersymmetric cases\footnote{ For that, it will be of great use
to employ superfield machinery such as the techniques of a
non-linear realization of supersymmetry \cite{bg,rt,gpr,bik} or
the superembedding formulation \cite{soro,hrs}.}.  We will not pursue this
issue here but will simply assume that such an extension exists. For 
this reason and others that will follow we mostly concentrate 
on supersymmetric cases below.

Since the discussions have been kept to the level of a low energy
effective action, there are various limitations to the claims one
can make based on the results. For example, the solution
(\ref{ss}) should not be viewed as to represent the entire stringy
configurations of closed strings including the massive modes. To
be able to make such a statement (or a similar one), one would
probably have to consider a BI type action in a background that
contains all the massive closed string fields, whose proper
discussion would require a full string theory or string field
theory. Rather one should view the procedure as a novel and
effective way of bringing out a closed  string as appropriate
degrees of freedom for a {\em massless} open string in the extreme
coupling.

However this interpretation still faces a criticism that the
action of our starting point, eq.(\ref{bi}), is incomplete because
it does not contain certain higher derivative terms and therefore
it could be used only for slowly varying configurations. Although
more complete resolution of this problem would have to wait until
the arrival of a superspace formulation of a D-brane action through
a partial breaking of supersymmetry, it might be useful to recall
that there could be be a field redefinition that removes some (but
not all) of the higher derivative terms. An example of a field
redefinition that removes certain higher derivative terms appeared
in the discussion of a 3-brane action in \cite{gpr}. Another
example is \cite{gkpr} where a field redefinition is introduced
for a comparison of a four dimensional super Yang-Mills action and
a BI action. After such a field redefinition, if necessary, the
resulting action might still allow a connection to a string
configuration. It is also relevant to note that the equation,
$\pa_\m {\cal K}^{\m\n}=0$, does not depend on the detailed form
of the action. It is the Bianchi identity of the dual
field\footnote{By the {\em dual} field we mean ${\cal K}$ or $G$
(Which of the two should be clear from the context), while the
original field refers to $F$. }, $G=*{\cal K}$, and the dual field
will satisfy the Bianchi identity irrespective of the detailed
form of the original action.

There is another issue worth discussing.
The picture seems to be contradictory to the fact that we have
open string {\em boundary conditions} to start with. In other
words, the open strings might remain as open strings even under
the strong coupling limit since their attachment to the branes are
realized as the boundary conditions. The resolution of this
puzzle might come from the fact that in general, the
boundary conditions must be consistent with the given background.
This apparently innocuous statement has not been much appreciated,
partially because we are more used to a flat background where no
moduli parameters take extreme values such as infinity. In such
backgrounds, one can impose Neumann or periodic boundary
conditions without any obstacle. However, a more general
background with some of its moduli parameters taking extreme
values, may restrict the choice of boundary conditions. After all,
boundary conditions themselves should be considered as a part of
the background and as such they should not contradict with the
rest of the data of the background. There is a familiar background
that can provide a concrete example: consider open strings in a
flat background with a constant B-field. The boundary
condition is
\be
g_{\m\n}\,\pa_nX^\n+2\pi\a'B_{\m\n}\,\pa_tX^\n \;\vert_{\mbox{bd}}=0
\ee
One can {\em not} take $B\rightarrow\infty$ imposing the Neumann boundary
condition at the same time. With these discussions we will assume that
the conversion is true at the level of the low energy effective
action. Furthermore we will assume that it will remain true at the
level of the full string discussion. We now turn to its implications.

Recently it has
been shown \cite{sst,gmms} that closed strings can be
decoupled from open strings in a background where the background
electric field approaches its critical value. One of the lessons of
these works is that the conventional lore that open strings need
closed strings is in fact a background-dependent statement.
One may take one step further and consider a general construction of
open-closed string theory with
open string fields only. (Closed strings {\em without} explicit closed
string fields was discussed in the past \cite{st,msrw,gt,siegel}.)
In other words, instead of putting closed  strings explicitly in the
kinematic setup one may start only with open
string world sheet Lagrangian.

The reason for considering such a kinematic setup is that it seems
better suited for the possible proof of the duality between a open
description and a closed string description. Let us start with a
open string theory with a very {\em small} coupling constant. The next
step is to apply S-duality. At the level of a low energy
effective action, it is a well known operation as we discussed in
the previous section.
After S-dualizing the system one can employ the argument that the
appropriate degrees of freedom are now those of a closed string.
Therefore, if true, the conversion will lead to very general
concept of duality between the two descriptions, (which in turn
reduces to low energy duality between field theories and gravity
theories). The dual closed string description will be {\em
strongly} coupled: the duality under consideration is different
from the familiar world sheet open-closed duality because the
latter is considered in the usual kinematic setup and furthermore
in the dual channel the description is still {\em weakly} coupled. It is
amusing to note that, as proposed in \cite{iyp}, this picture
implies that the geometry in which the dual closed string
propagates is the same\footnote{The relevance of curved
backgrounds for a BI action was discussed in
\cite{dt,flz,iyp,pst,dt2} in connection with AdS/CFT.} (but with
strong coupling ) as the one for the original open string. (See
Fig. 1.)
\begin{figure}[!ht]
\centerline{
        \begin{minipage}[b]{8cm}
               \epsfxsize=8cm
                \epsfbox{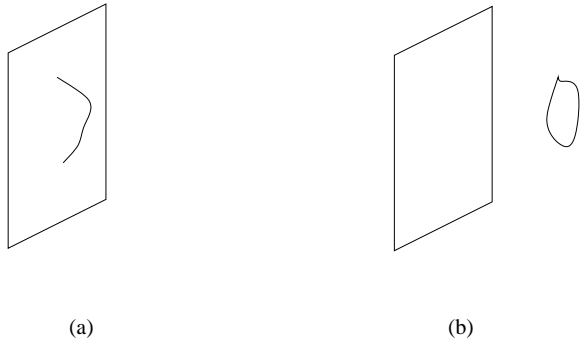}
        \end{minipage}
}
\caption{(a) open string description
(b) closed string description }
\label{fig:2all5}
\end{figure}

We are ready to discuss how the conversion ``derives'' AdS/CFT
duality. AdS/CFT can be motivated by taking a viewpoint
that there are two different but dual stringy descriptions of the
same objects, D-branes. One description is via open
strings with mixed Dirichlet and Neumann boundary conditions. In
the other description one considers type IIB closed string
theory expanded around the D-brane soliton solutions.\footnote{As 
discussed above, it should not be confused
with the familiar channel duality. In \cite{iyp} it was called
fundamental-solitonic duality to avoid the confusion.} Upon taking
a low energy near horizon limit, the viewpoint leads, e.g., in the
case of D3 branes, to the duality between $\N=4, D=4$ SYM theory
and $\N=8, D=5$ gauged supergravity. To make the discussion
slightly more general, consider a open string attached to an
arbitrary odd dimensional brane. The open string propagates in the
curved background produced by the presence of the brane. We start
with a weakly coupled opens string theory in a pure open string
formulation.\footnote{ Or one may consider a decoupling limit
where the asymptotic close strings decouple. The limit results
from a slight modification of the familiar scaling limits of the
moduli parameters. Consider open strings
 with an  extremely weak coupling.
With a vanishing coupling constant the asymptotic closed strings
will be decoupled. One then takes a large N limit such that $g_sN$
becomes fixed. However, $g_sN$ is taken to be {\em small} to avoid
suppressing massive stringy excitations. The condition, $g_s\ra
0$, suppresses the dynamical generation of close strings, which
otherwise would propagate off the branes. However the open string
theory is still an interacting theory on the world volume since
the non-vanishing value of $g_sN$, which is the effective coupling
of the open strings on the world volume. } and go to a strong
coupling limit by S-duality. After the duality the appropriate
degrees of freedom are a closed string in the same background. The
closed string should be of type IIB and strongly coupled. In case
of a D3 brane, the resulting closed string is in the D3 brane
background, but can be considered as {\em weakly} coupled due to the SL(2,Z)
self-duality \cite{schwarz}: the weakly coupled IIB closed string is 
connected by a chain of dualities to the starting point, a weakly
coupled open string. In the low energy this leads to the duality
between ${\cal N}=4$ SYM theory and IIB supergravity on AdS$_5\times $S$^5$.
Similarly, starting with an open string in the background of 
D2/D4\footnote{The relevance of the world volume theories of D2/D4 branes
was discussed in \cite{m}.} branes we will get AdS/CFT conjecture
concerning AdS$_{4/7}\times $S$^{7/4}$.

One can give similar arguments for matrix theories. Consider an
open superstring with the fermionic coordinates of opposite
chiralities. Take fully Neumann boundary conditions. Apply the
conversion procedure, i.e., one start with a very weak coupling
and consider S-duality to go to a strong coupling limit. The
closed string that appears will be of type IIA. In particular it
will be strongly coupled. The strong coupling limit of IIA is
M-theory \cite{w2}. On the other hand we can T-dualize the original open
string to an open string with Dirichlet boundary conditions for
the nine space dimensions. Therefore, one end of this operation is
an open string with Dirichlet boundary conditions and the other
end is M-theory. Once we go to a low energy limit, the two
theories respectively reduce to quantum mechanics and eleven
dimensional supergravity:  we have the matrix theory conjecture of
M-theory \cite{bfss}. On the other hand, if we start with an opens string
whose fermionic coordinates have the {\em same} chiralities, we
will get IKKT matrix theory conjecture \cite{ikkt}.
\begin{figure}[!ht]
\centerline{
        \begin{minipage}[b]{15cm}
               \epsfxsize=15cm
                \epsfbox{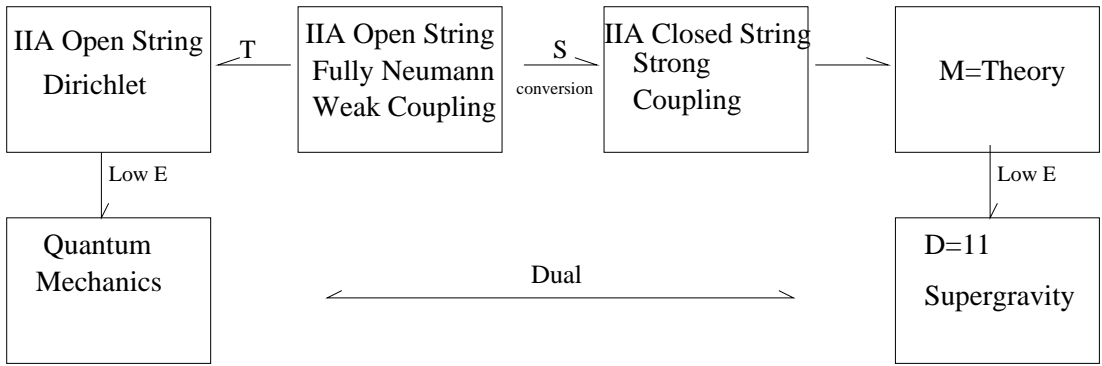}
        \end{minipage}
} \caption{Duality chain for BFSS matrix theory }
\label{fig:2all5}
\end{figure}

\section{Conclusion}

AdS/CFT conjecture was motivated in \cite{iyp} by starting from an
axiomatic assumption that there are two dual descriptions of D-branes, 
open and closed. We have argued here that the duality may originate
from the conversion of an open strong into a closed string under a
large coupling limit where the coupling constant approaches infinity.
For that, we considered a ({\em non}-tachyonic) generalization of the 
analysis of \cite{ghy} to a curved background with a constant $B$-field.
Although we studied only a bosonic Born-Infeld action we believe that
a similar analysis should be possible for a supersymmetric case, which
is an interesting open problem. 
We noted that the duality under consideration implies, in low energy limits,
very general dual relations between gauge theories and gravity theories. In
particular it seems that AdS/CFT and matrix theory conjectures come
from the same root, the conversion of an open string into a closed string
under a strong coupling limit. It will be interesting to promote the 
discussions to a full string theory analysis.\\

\ni{\bf Note Added}: After this work was published, a paper \cite{kk}
appeared which has partially related discussions.

\vspace{1in}
\noindent {\Large \bf Acknowledgements}\\
For their valuable discussions I would like to thank B. Craps,
N.S. Deger, A. Kaya, P. Kraus, F. Larsen, M. Ro\v{c}ek, I. Rudychev,
W. Siegel, T.A. Tran, A.A. Tseytlin, P. Yi and especially
C.N. Pope. This work is supported by US
Department of Energy under grant DE-FG03-95ER40917.

\section*{Appendix}
In section 2 we imposed $g_{\cc i}=0 $ for simplicity. Here we
relax the condition. Since the calculations for the $B$-part
remain the same we will not consider them. With $g_{\cc i} \neq 0$
the determinant can be rewritten as
\be -\mbox{Det}(g_{\m\n}+F_{\m\n}  ) = -g_{\cc \cc} \mbox{Det}
(h_{ik}) - \Ep_i D^{ik} \Em_k \ee
where
\bea {E^{\pm}}_i &=& E_i\pm g_{\cc i} \;\;\;\;,\;\;\;\;
                         h_{ij} = g_{ij}+F_{ij} \nn\\
D^{ij}\;\; &=& (-1)^{i+j}\D^{ji}(h) \;\;\;,\;\;\;
D=\mbox{Det}(h)\,  h^{-1}
 \eea
where $\D$ represents the minor matrix of $h$. The canonical
momenta are
\be \pi^i =\fr{\t }{\sqrt{-\mbox{Det}(g_{\m\n}+F_{\m\n})}}\;
             \fr{1}{2} ( \Ep_k D^{ki}+  D^{ki}\Em_k )
\ee
After some algebra one can show that
\bea {\cal H}
                &=&   \fr{\t}{\sqrt{-\mbox{Det}(g_{\m\n}+F_{\m\n} )} }
                      \left( -\mbox{Det}(h) g_{\cc \cc}
                   +\fr{1}{2}g_{\cc i} \left(\Ep_kD^{ki}-D^{ik}\Em_k  \right)
                                              \right)
\eea
The Hamiltonian in terms of the canonical variables can be obtained by
solving the following equation that it satisfies,
\bea
 {\cal H}^2-2g_{0l} u^{li} F_{ij}\p^j \;{\cal H}
 + \fr{1}{G^{\cc \cc}} \left( \pi^i g_{ij} \pi^j\!
 +(F_{ij}\pi^j)^2 \right) =0\eea
where $(\!F_{ij}\pi^j)^2 \equiv g^{ik} F_{ij}\pi^j F_{km}\p^m$ and
$\t$ has been set to zero. We have also introduced, $u^{ij}$, the
inverse matrix of $g_{ij}$. In general, $u^{ij}$ is {\em not} \,the
same as $g^{ij}$, the (ij)-th component of $g^{\m\n}$ although
that was true in the case we considered in section two, i.e.,
in the case of $g_{i0}=0$. As before we perform a Legendre
transformation with
\bea K^{ij}=-\;\fr{\; g_{0l} u^{li} \p^j - g_{0l} u^{lj} \pi^i
+\fr{2}{G^{\cc \cc}}( F^i_l\pi^l \pi^j- F^j_l\pi^l \pi^i   )  } {
2{\cal H}-2g_{0l} u^{li} F_{ij}\p^j }\eea
where $F^i_l \equiv G^{ik}F_{kl}$. It is straightforward to show
\bea {\cal L}'= - \fr{2}{G^{\cc \cc}}\;\fr{\pi^i g_{ij} \pi^j}
               {\;2{\cal H} -2g_{\cc l} u^{li} F_{ij}\pi^j   }
               \eea
Using the same definitions for the components of ${\cal K}^{\m\n}$, i.e.,
\bea {\cal K}^{\cc i} = -\pi^i\;\;\;,\;\;\; {\cal K}^{ij} &=&
K^{ij} \eea
One can show, after rather lengthy algebra, that 
\be \fr{1}{2}{\cal K}^{\m\n}{\cal K}_{\m\n}
                         =- \left(\fr{2}{G^{\cc\cc}}\right)^2
                         \;\fr{(\pi^i g_{ij} \pi^j)^2}
               {\;(2{\cal H} -2g_{\cc l} u^{li} F_{ij}\pi^j )^2  }
\ee
Therefore the dual Lagrangian has the same form as
before,
\be {\cal L}'=\sqrt{-\fr12{\cal K}^{^{\m\n}}{\cal K}_{_{\m\n}}}
\ee
and the discussions below equation (\ref{Kappa}) of section two remain 
the same.

\end{document}